\documentclass[aps,prb,twocolumn,superscriptaddress]{revtex4-2}
\usepackage{bm}
\usepackage{graphicx}
\usepackage[usenames]{color}
\usepackage{amsmath,amsfonts,amssymb}
\usepackage{hyperref}
\usepackage{ulem}
\usepackage{xcolor}
\usepackage{comment}

\newcommand{\ket}[1]{|{#1}\rangle}

\newcommand{\pa}{\partial}

\allowdisplaybreaks[4]

\newcommand{\tsukubaD}{Department of Physics, University of Tsukuba, Tsukuba, Ibaraki 305-8571, Japan}
\newcommand{\tsukubaG}{Graduate School of Pure and Applied Sciences, University of Tsukuba, Tsukuba, Ibaraki 305-8571, Japan}
\newcommand{\penn}{Department of Physics, 104 Davey Lab, The Pennsylvania State University, University Park, Pennsylvania 16802, USA}

\usepackage{version}	
\begin{document}
\title{Adiabatic Continuity of the Spinful Quantum Hall States}

\author{Koji Kudo}
\affiliation{\penn}
\affiliation{\tsukubaD}
\author{Yasuhiro Hatsugai}
\affiliation{\tsukubaD}
\affiliation{\tsukubaG}

\date{\today}

\begin{abstract}
 By using the extended Hubbard model of anyons, we numerically demonstrate
 the adiabatic deformation of the spinful quantum Hall (QH) states 
 by transmutation of statistical fluxes. While the ground state is always 
 spin-polarized in a series of $\nu=1$ integer QH system, the adiabatic 
 continuity between the singlet QH states at $\nu=2$ and $\nu=2/5$ is 
 confirmed. These results are consistent with the composite fermion theory with
 spin. The many-body Chern number of the ground state multiplet works as an 
 adiabatic invariant and also explains the wild change of the topological 
 degeneracy during the evolution. The generalized St\v{r}eda formula of spinful
 systems is justified.
\end{abstract}

\maketitle

\section{Introduction}
In these decades, topology has been coming to the fore in condensed matter
physics. The integer quantum Hall (IQH) 
effect~\cite{Klitzing_IQH_PRL80,Laughlin_PRL81}
is a prototypical example of
topologically nontrivial phase, where topological nature of the Chern number 
is the origin of the quantization of the Hall 
conductance~\cite{Thouless_TKNN_PRL82,Kohmoto_Chern_Ann85}.
Topological invariants also work as order parameters beyond the Ginzburg-Landau
theory based on the breaking symmetry, which demonstrates how topology brings 
further diversity to phases of matter. The electron-electron interaction gives 
even more enriched topological phenomena. The fractional quantum Hall (FQH) 
effect~\cite{Tsui_FQH_PRL82,Laughlin_FQH_PRL83} is topologically 
ordered~\cite{Wen_TO_PRB89,Wen_TO_AP95} and hosts fractionalized excitations 
carrying the fractional charge and fractional 
statistics~\cite{Wilczek_anyon1_PRL82,Wilczek_anyon2_PRL82,Arovas_FS_PRL84,Haldane_boson_PRL83,Halperin_FS_PRL84}.
Even though the origin of the energy gap is intrinsically different in the IQH 
and the FQH effects, the composite fermion
theory~\cite{Jain_CFT_PRL89,jain_2007}
enables us to understand their underlying physics in a unified scheme:
the FQH state at the filling factor $\nu=p/(2mp\pm1)$ with $p,m$ integers is
interpreted as the $\nu=p$ IQH state of composite fermions carrying $2m$ 
fluxes. Their adiabatic continuity by trading the external fluxes for the 
statistical ones has been demonstrated in various
situations~\cite{Greiter_NPB90,Greiter_NPB92,Kudo_PRB20,Pu_anyon_PRB21,Greiter_AH_PRB21,Kudo_BEC_PRB21},
which justifies validity of the composite fermion picture. 
On a torus, the many-body Chern number~\cite{Niu_NTW_PRB85} remains constant 
during the adiabatic evolution. This describes the wild change 
of the topological degeneracy in a similar form of the St\v{r}eda 
formula~\cite{Streda_IOP82}, which we call the generalized St\v{r}eda 
formula~\cite{Kudo_PRB20}.

The internal degree of freedom generates further diversity in the FQH effects.
A typical system is the spinful FQH systems where small Zeeman splitting is neglected~\cite{Halperin_lmn_Helv83,Clark_experiment1_PRL89,Eisenstein_experiment2_PRL89,Chakraborty_review_Sur90,Sondhi_PRB93,Wu_CFspin_PRL93}. 
Multilayer
systems~\cite{Suen_bi_PRL92,Eisenstein_bi_PRL92,He_bi_93,Scarola_CF_PRB01,Eisenstein_14}, and (multilayer) 
graphene~\cite{Du_Nature09,Bolotin_Nature_09,Nomura_PRL06,Apalkov_PRL06,Toke_PRB06,Toke_PRB07,Hamamoto_PRB12,Balram_PRB15,Wu_Nano17,Zibrov_NatureP18,Kudo_JPSJ18,Faugno_PRB20,Abouelkomsan_PRL20,Ledwith_PRR20,Repellin_PRR20,Wilhelm_PRB21,Xie_Nature21}
also give exotic FQH states that cannot be observed in single component 
systems. They are not only fundamentally interesting in its own right but may 
also provide a platform for topological quantum computation 
based on the non-Abelian bradings~\cite{Kitaev_Annals03,Nayak_RMP08}, which has
attracted a great interest for these decades. The composite fermion theory is 
remarkably useful in the multicomponent FQH systems as well~\cite{jain_2007}. 
For example, the spin structure of the FQH states in the limit of
vanishing Zeeman energy depends strongly on the filling factor. This selection 
rule for the spin can be predicted by the corresponding IQH state of composite 
fermions~\cite{Wu_CFspin_PRL93}. This can be applied to other degree of freedom
such as a layer index~\cite{Scarola_CF_PRB01} and the valley 
degree~\cite{Toke_PRB06}. The main goal in this work is to demonstrate their
adiabatic continuity and reveal the topological properties during the evolution
of the flux-attachment.

We below numerically analyze the extended Hubbard model of two-component 
anyons. They are spinful anyons but it can be applied to other 
degree of freedom such as the layer index. We demonstrate that the spin-singlet
IQH state at $\nu=2$ is adiabatically connected to the $\nu=2/5$ singlet FQH
state~\cite{Halperin_lmn_Helv83} while the topological degeneracy changes 
wildly. The adiabatic continuity between the bosonic IQH state at 
$\nu=2$~\cite{Senthil_PRL13,Furukawa_PRL13} and the singlet FQH state at 
$\nu=2/3$~\cite{Wu_CFspin_PRL93} is also confirmed. On the other hand, a 
series of $\nu=1$ IQH system always gives maximally spin-polarized ground 
states. These results are
consistent with the composite fermion theory with spin~\cite{Wu_CFspin_PRL93}.
We also confirm that the many-body Chern number of the 
ground state remains constant during the adiabatic evolution and also describes
the wild change of the topological degeneracy. This justifies validity of the 
generalized St\v{r}eda formula~\cite{Kudo_PRB20} in spinful QH 
systems.

\section{Extended Anyon-Hubbard model}
\subsection{Spinful anyons}
Let us consider a toroidal system of anyons on a square lattice under the
magnetic field. Anyons are with two components labeled by spin with $S=1/2$ 
but it can be applied to other degree of freedom such as the layer index. 
Counterclockwise exchanges of particles with same spin give the phase factor 
$e^{i\theta}$. Although an exchange of opposite spins gives a different 
many-body state, we assume that 
the two successive operations give the phase factor $e^{i2\theta}$. Namely, a 
local move around another always gives $e^{i2\theta}$ irrespective of their 
spins, which is a physically natural extension of the fractional statistics
to two-component systems~\cite{Hosotani_PRB90,Lee_PRL90}.
\subsection{Hamiltonian}
Modeling the anyons as fermions with the statistical fluxes, we define the 
Hamiltonian $H=H_\text{kin}+H_\text{int}$ with
\begin{align}
 &H_\text{kin}=-t\sum_{\alpha,\langle ij\rangle}
 c_{i\alpha}^\dagger e^{i\phi_{ij}}e^{i\theta_{ij}}c_{j\alpha},
 \label{eq:Hkin}\\
 &H_\text{int}=
 U\sum_{i}n_{i\uparrow}n_{i\downarrow}
 +V\sum_{\langle ij\rangle}n_{i}n_{j},
\end{align}
where $c^\dagger_{i\alpha}$ is the creation operator for a fermion~\cite{fermi}
with spin $\alpha=\uparrow,\downarrow$ on site $i$,
$n_{i\alpha}\equiv c^\dagger_{i\alpha}c_{i\alpha}$,
$n_i\equiv n_{i\uparrow}+n_{i\downarrow}$, $\langle ij\rangle$ indicates the 
summation over the nearest-neighbor pairs of sites, and $e^{i\phi_{ij}}$ 
describes the external magnetic field~\cite{Hatsugai_PRL99}. The statistical 
fluxes are introduced by $e^{i\theta_{ij}}$, see below for details. Unless 
$e^{i2\theta}=1$, particles carry fractional 
fluxes, which implies that $\theta_{ij}$ is ill defined if two or more particle
coordinates coincide. To avoid the singularities, we set $U=+\infty$ that 
results in the hard-core constraint $c^\dagger_{i\alpha}c^\dagger_{i\beta}=0$ 
for any spin $\alpha,\beta$. The other parameters are set as $t=1$ and 
$V\geq0$. 

Our system preserves SU(2) spin-rotational symmetry since the Hamiltonian is 
expressed as
\begin{align}
 H=\sum_{ij}\left(t_{ij}\bm{c}^\dagger_i\bm{c}_j+V_{ij}n_in_j\right),
\end{align}
where $\bm{c}^\dagger_i=(c^\dagger_{i\uparrow},c^\dagger_{i\downarrow})$, 
$t_{ij}$ is a function of the operators $n_k$'s, and $V_{ij}$ is a constant. 
This is obviously invariant under the transformation 
$\bm{c}^\dagger_i\rightarrow\bm{c}^\dagger_iu$ with $u\in\text{SU}$(2). 

\subsection{Statistical fluxes with spin}
Let us mention how to define the gauge field $\theta_{ij}$ in 
Eq.~\eqref{eq:Hkin}. Following the method in 
Ref.~\onlinecite{Wen_anyon_PRB90,Hatsugai_cylinder_PRB91,Hatsugai_torus_PRB91},
we first construct the hopping Hamiltonian for spinless anyons under with 
the statistical phase $\theta$ under the magnetic field as 
$H'_\text{kin}=-t\sum_{\langle ij\rangle}c_{i}^\dagger e^{i\phi_{ij}}e^{i\theta_{ij}'}c_{j}$, 
where $\theta_{ij}'=\sum_{k\neq i,j}A_{ijk}c_{k}^\dagger c_{k}$ with $A_{ijk}$ 
real. The boundary conditions are modified to ensure the braid group on a torus
as described in the next paragraph. Within this framework, we then define the 
Hamiltonian in Eq.~\eqref{eq:Hkin} with 
\begin{align}
 \theta_{ij}
 =\sum_{k\neq i,j}A_{ijk}
 (c_{k\uparrow}^\dagger c_{k\uparrow}+c_{k\downarrow}^\dagger c_{k\downarrow})
 =\sum_{k\neq i,j}A_{ijk}n_k
\end{align}
under the same 
boundary conditions. Hoppings in $H'_\text{kin}$ properly give the phase 
factors $e^{i\theta}$ and $e^{i2\theta}$ for particle exchanges and for moves 
of particles around another, respectively. In the same manner, hoppings in 
$H_\text{kin}$ give $e^{i\theta}$ and $e^{i2\theta}$ for particle exchanges 
within same spin and for moves around another particle irrespective their spins
as well.

A particle exchange performed by the global moves on a torus requires 
modification of the boundary conditions for spinless 
anyons~\cite{Einarsson_PRL90,Wen_anyon_PRB90,Hatsugai_cylinder_PRB91,Hatsugai_torus_PRB91}. Accordingly, the Hilbert space of anyons with $\theta/\pi=n/m$ 
($n,m$: coprimes) is spanned by the basis $\ket{\{\bm{r}_k\};w}$, where 
$\{\bm{r}_k\}$ is the particle configuration and $w=1,\ldots,m$ is the 
additional label associated with the boundaries: when a 
particle crosses the boundary in the $x$ ($y$) direction, the label is shifted 
from $w$ to $w-1$ (the phase factor $e^{iw\theta}$ is given).
This realizes non-local nature of anyons, which we employ in our spinful 
system in the same way. This condition does not break the SU(2) symmetry.

The existence of the label $w$ implies that $\text{dim}\,H$ with 
$\theta/\pi=n/m$ is
$m$ times larger than that with fermions or bosons even for the same particle 
and the same site numbers, meaning that $H$ changes discretely as 
$\theta$ is changed continuously. Nevertheless, as shown below, the energy gaps
of the QH states behave smoothly in the evolution of the 
flux-attachment although the ground state degeneracy is wildly changed. Using 
this smoothness found in a dense set of the energy gaps, we define ``adiabatic 
continuity''.

\section{Adiabatic continuity}
By the above setup, we investigate the adiabatic continuity of the 
spinful QH states under the flux-attachment 
transformation~\cite{Jain_CFT_PRL89,Greiter_NPB90,Wu_CFspin_PRL93}. This 
transformation trades the external magnetic fluxes for the statistical fluxes
while remaining their total number constant, i.e.,
\begin{align}
 N_\phi+N_p\frac{\theta}{\pi}=\text{const},
\end{align}
where $N_\phi$ is the number of external fluxes and 
$N_p=N_\uparrow+N_\downarrow$ is the particle number. This implies that a 
fermionic system at $\nu\equiv N_p/N_\phi=p$ is transformed to systems of 
anyons with the statistical angle $\theta$ at
\begin{align}
 \nu=\frac{p}{p(1-\theta/\pi)+1}.
 \label{eq:AH}
\end{align}
We call such a set of transformed systems the family of 
the $\nu=p$ IQH system. The following discussions focus on the most basic 
cases, $p=1$ and $2$.

\subsection{Family of the $\nu=1$ IQH system}
\subsubsection{Energy gap}
Let us first consider a family of the $\nu=1$ IQH system. We numerically
demonstrate that the lattice analogue of the Halperin $lll$ 
state~\cite{Halperin_lmn_Helv83} emerges at $\nu=1/l$ with $l=1,2,3$ and 
intermediate systems of anyons also gives maximally polarized ground state.
This result justifies the composite fermion theory with 
spin~\cite{Wu_CFspin_PRL93}.

\begin{figure}[t!]
 \begin{center}  
  \includegraphics[width=\columnwidth]{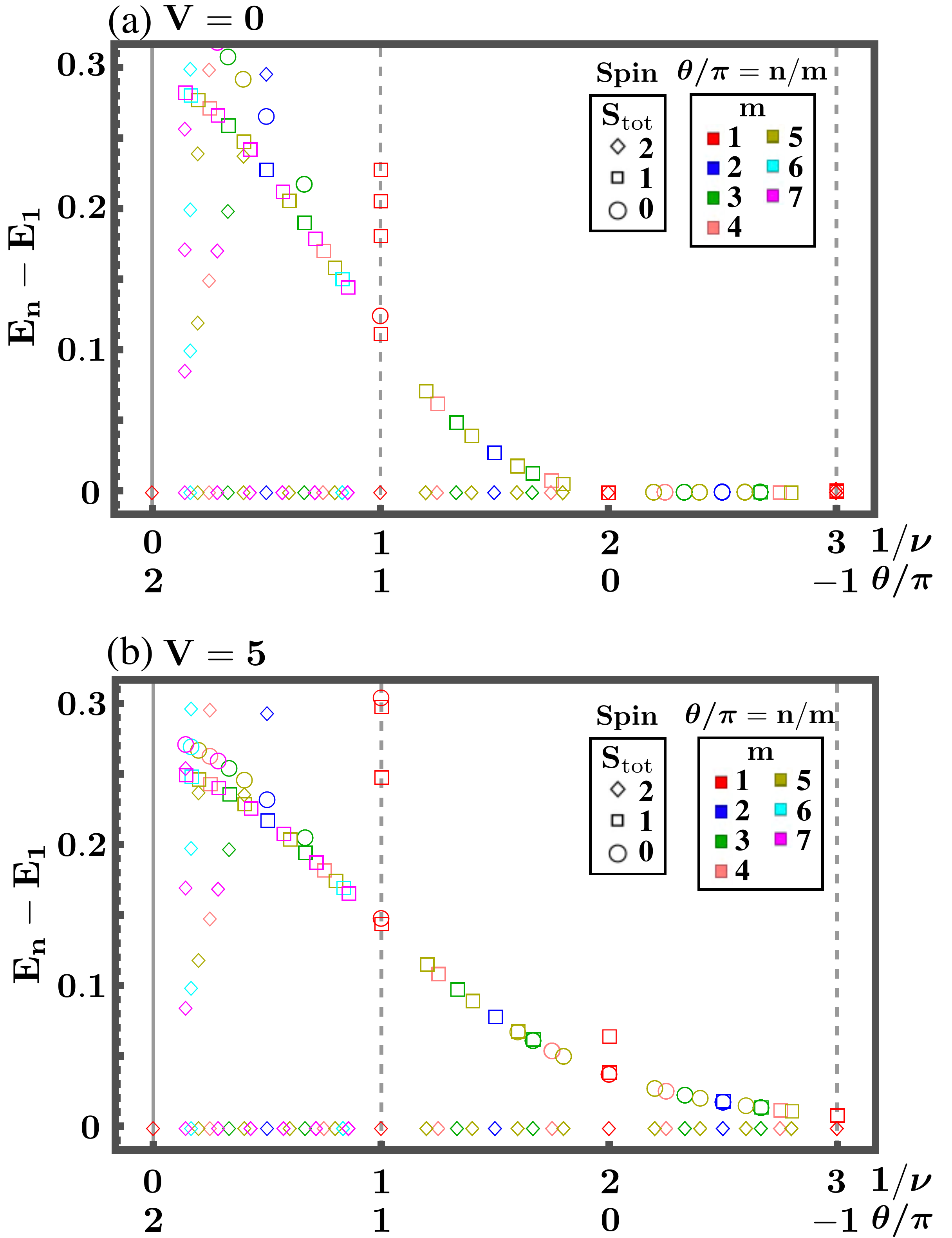}
 \end{center}
 \caption{
 Energy gaps as functions of $1/\nu$ for (a) $V=0$ and (b) $V=5$. We set 
 $N_p=4$ and $N_x\times N_y=9\times9$. We plot the lowest 15 energies within
 the $S_z^\text{tot}=0$ sector at each $1/\nu$. The vertical dashed lines 
 represent fermionic systems.
 }
 \label{fig:nu1_1}
\end{figure}
In Fig.~\ref{fig:nu1_1}(a), we plot the energy gap as functions of 
$1/\nu$, setting $N_p=4$, $V=0$, and the system size as 
$N_x\times N_y=9\times9$. Here, $\theta/\pi$ changes under the constraint in 
Eq.~\eqref{eq:AH} with $p=1$. The SU(2) spin-rotational symmetry allows us to 
label the eigenstates with total spin $S_\text{tot}$~\cite{calS}. At $\nu=1$, 
we obtain the maximally polarized IQH state with 
$S_\text{tot}=S_\text{tot}^\text{max}=2$, which is the lattice analogue of the 
Halperin 111 state. Even though $\nu$ is integer, the Hubbard interaction is 
crucial here since the lowest Landau level (LLL) is partially filled at $\nu=1$
(the LLL has $2N_\phi$ single-particle states while the particle number is 
$N_p=N_\phi$). Following the argument of the flat band 
ferromagnetism~\cite{Mielke_FB1_91,Mielke_FB2_91,Tasaki_PRL92,Tasaki_PTP98},
one expects the spin-polarized ground state. The first excited state gives 
$S=S_\text{tot}^\text{max}-1$, which is consistent with the spin wave
of the polarized QH states. This means that the obtained finite gap is a 
finite size effect, but it survives as $1/\nu$ increases and then closes at 
$\nu=1/2$. This suggests that the spin-polarization at the fractional fillings 
is understood by the maximally polarized IQH state~\cite{Wu_CFspin_PRL93}.

The gap closing at $\nu=1/2$ is explained by the composite fermion 
theory~\cite{Jain_CFT_PRL89,Wu_CFspin_PRL93}.
Noting $U=\infty$ and $V=0$ on a lattice, let us consider an interaction 
$\sum_{ij}\delta^2(z_i-w_j)$ in the continuum system in the disk geometry, 
where $z_j=x_j-iy_j$ and $w_j=x_j-iy_j$ are the positions of particles with 
$\alpha=\uparrow$ and $\downarrow$, respectively. Within the LLL, this 
interaction gives the zero-energy degenerate eigenfunctions of bosons at 
$\nu=1/2$:
\begin{align}
 \Psi_{\nu=\frac{1}{2}}^S
=\prod_{i<j}(z_i-z_j)\prod_{i<j}(w_i-w_j)\prod_{i,j}(z_i-w_j)\Phi_{\nu=1}^S.
\end{align}
Here, $\Phi_{\nu=1}^S$ is a LLL projected state at $\nu=1$ with total spin $S$,
which is macroscopically degenerate for general $S$ as the LLL is partially 
filled while $\Phi_{\nu=1}^{S=S_\text{max}}$ is unique. The gap 
closing at $\nu=1/2$ in Fig.~\ref{fig:nu1_1}(a) is consistent with this fact.

The discussion implies that the adiabatic continuity in a wider range of
$\nu$ is established by turning on the nearest-neighbor interaction $V$. 
Figure~\ref{fig:nu1_1}(b) is the same as Fig.~\ref{fig:nu1_1}(a) but for $V=5$.
The gap at $\nu=1/2$ becomes finite and the $\nu=1$ IQH state is 
adiabatically connected to the the lattice analogue of the Laughlin state 
(Halperin 333 state) at $\nu=1/3$. 

\subsubsection{Topological degeneracy and Chern number}
\begin{figure}[t!]
 \begin{center}  
  \includegraphics[width=\columnwidth]{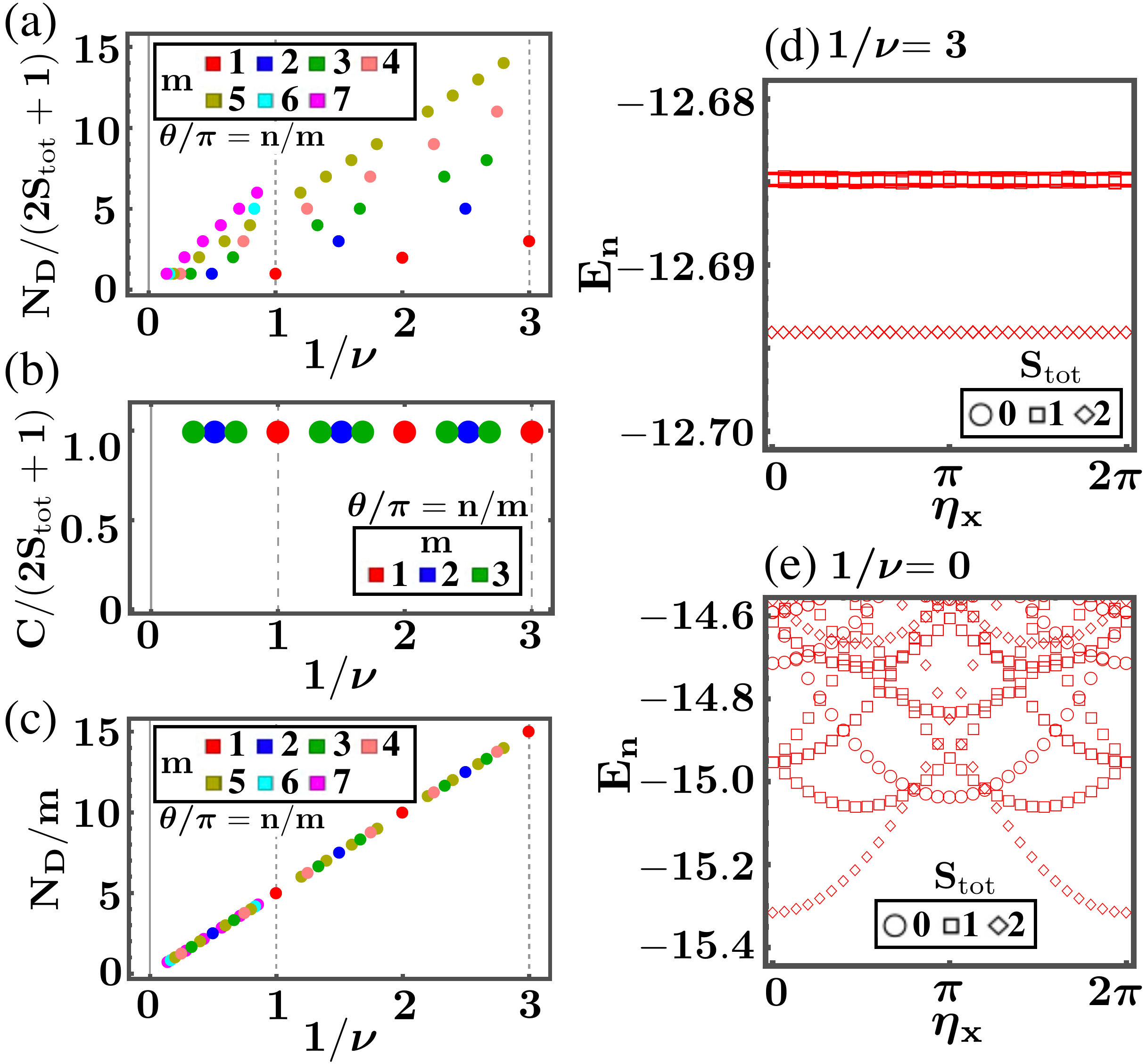}
 \end{center}
 \caption{
 (a) Ground state degeneracy $N_D$ divided by the spin degeneracy 
 $2S_\text{tot}+1$ ($=5$ in this case) as a function of $1/\nu$.
 (b) Many-body Chern number.
 (c) Ground state degeneracy $N_D$ divided by the denominator of $\theta/\pi$. 
 (d)(e) Spectral flows at (d) $1/\nu=3$ of fermions and (e) 
 $1/\nu=0$ of bosons.
 In the all panels, the same setting as Fig.~\ref{fig:nu1_1}(b) is used.
 }
 \label{fig:nu1_2}
\end{figure}
Because of the spin-polarization, topological properties of the obtained 
ground state are identical to that of 
spinless systems~\cite{Kudo_PRB20} except for the $(2S_\text{tot}^\text{max}+1)$-fold spin degeneracy. 
Assuming that the states are degenerate if their energy difference is less than
0.001 in Fig.~\ref{fig:nu1_1}(b), we plot in Fig.~\ref{fig:nu1_2}(a) the ground
state degeneracy $N_D$. It changes wildly even though the energy 
gap behaves smoothly. In Fig.~\ref{fig:nu1_2}(b), imposing the twisted boundary
conditions, we compute the many-body Chern number~\cite{Niu_NTW_PRB85},
\begin{align}
 C=\frac{1}{2\pi i}\int_{T^2}d^2\eta F,
\end{align}
where $\vec{\eta}=(\eta_x,\eta_y)$ is the twisted angles, 
$T^2=[0,2\pi]\times[0,2\pi]$, $F=(\pa A_y/\pa\eta_x)-(\pa A_x/\pa\eta_y)$, 
$A_{x(y)}=\text{Tr}\,[\Phi^\dagger(\pa\Phi/\pa\eta_{x(y)})]$, and 
$\Phi(\vec{\eta})=(\ket{G_1(\vec{\eta})},\cdots,\ket{G_{N_D}(\vec{\eta})})$
is the ground state multiplet. As shown in Fig.~\ref{fig:nu1_2}(b), the Chern
number $C$ works as an adiabatic invariant numerically. 

Extending the generalized St\v{r}eda formula for spinless 
anyons~\cite{Kudo_PRB20}, the wild change of the degeneracy is described by the
many-body Chern number as 
\begin{align}
 \frac{\Delta N_D}{\Delta(m/\nu)}=C,
 \label{eq:GSF}
\end{align}
where $m$ is the denominator of $\theta/\pi$ and $\Delta$ represents the 
difference for two possible cases in a family. This works even for polarized 
states with the spin degeneracy because the additional factors appear in both 
sides of Eq.~\eqref{eq:GSF} as $N_D\rightarrow(2S_\text{tot}^\text{max}+1)N_D$ 
and $C\rightarrow(2S_\text{tot}^\text{max}+1)C$. 
In Fig.~\ref{fig:nu1_2}(c), we plot $N_D/m$ as a 
function of $1/\nu$. The slope is surely identical to $C$ $(=5$ in this case),
which is consistent with Eq.~\eqref{eq:GSF}.

Let us mention the twisted boundary conditions. In our model, they are 
defined as follows: when an anyon hops across the boundary in $x$ ($y$) 
direction, the phase factor $e^{i\eta_x\delta_{w1}}$ ($e^{i\eta_y}$) is given 
to the basis $\ket{\{\bm{r}_{k\uparrow}\},\{\bm{r}_{k\downarrow}\};w}$. 
In Figs.~\ref{fig:nu1_2}(d) and 
\ref{fig:nu1_2}(e), we plot the energies at $1/\nu=3$ and $1/\nu=0$ of
Fig.~\ref{fig:nu1_1}(b) as functions of $\eta_x$ fixing $\eta_y=0$. While the 
FQH state at $\nu=1/3$ is insensitive to the boundary condition, the spectral 
flow at $1/\nu=0$ gives the strong $\eta_x$-dependence and the gap is closed. 
This is consistent with the emergence of Nambu-Goldston modes of bosonic
superconductor~\cite{Zhang_Boson_B92}.

\subsection{Family of the $\nu=2$ IQH system}
\subsubsection{Energy gap}
Let us next consider a family of the $\nu=2$ IQH system. We demonstrate
the adiabatic continuity between the singlet FQH states at $\nu=2$ and 
$\nu=2/5$, which correspond to the Halperin 110 state and Halperin 332 state, 
respectively. It is also shown that the $\nu=-2$ bosonic IQH 
state~\cite{Senthil_PRL13,Furukawa_PRL13}, a symmetry-protected topological 
phase of bosons discussed in 
Ref.~\onlinecite{Chen_Science12,Chen_PRB13,Lu_PRB12}, is adiabatically 
connected to the singlet FQH state at $\nu=-2/3$~\cite{Wu_CFspin_PRL93}.
These results are consistent with the composite fermion theory with 
spin~\cite{Wu_CFspin_PRL93}.

\begin{figure}[t!]
 \begin{center}  
  \includegraphics[width=\columnwidth]{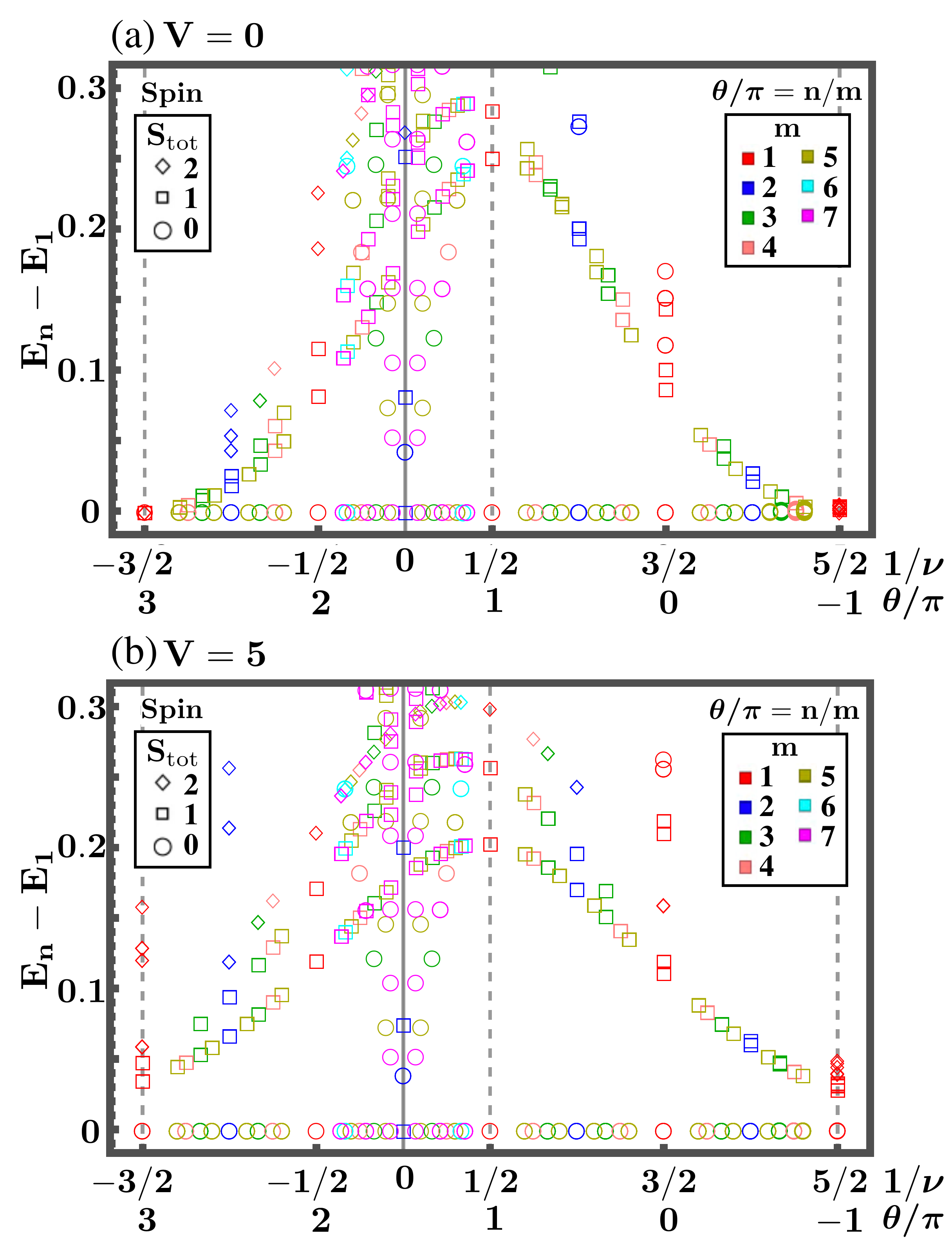}
 \end{center}
 \caption{
 Energy gaps as functions of $1/\nu$ for (a) $V=0$ and (b) $V=5$. We set 
 $N_p=4$ and $N_x\times N_y=8\times8$. We plot the lowest 25 energies within
 the $S_z^\text{tot}=0$ sector at each $1/\nu$. The vertical dashed lines 
 represents fermionic systems.
 }
 \label{fig:nu2_1}
\end{figure}
We plot the energy gap as functions of $1/\nu$ in 
Fig.~\ref{fig:nu2_1}(a), setting $N_p=4$, $V=0$, and $N_x\times N_y=8\times8$.
At $\nu=2$, we obtain the 
spin-singlet IQH state with $S_\text{tot}=S_\text{tot}^\text{min}\equiv0$.
If the Hubbard $U$ vanishes, the ground state is a completely occupied
LLL (Halperin 110 state). The obtained ground state is expected to be
topologically equivalent to that even for the infinite Hubbard $U$ since the 
density per site is very small. The energy gap at $\nu=2$ survives as 
$1/\nu$ increases. It closes at $\nu=2/5$ since maximally polarized states 
at least can be ground states in the fermionic system with $U=\infty$. 
Figure~\ref{fig:nu2_1}(b) is 
the same as Fig.~\ref{fig:nu2_1}(a) but for $V=5$. This indicates that the 
nearest neighbor interaction brings the singlet ground state at $\nu=2/5$, and
consequently the adiabatic continuity between the $\nu=2$ and $\nu=2/5$ is 
established. The ground state at $\nu=2/5$ is 5-fold degenerate and its 
many-body Chern number $C$ is 2, which is consistent with the Halperin 332 
state.

In a family of $\nu=2$ IQH system, two systems at $+\nu$ and $-\nu$ are not 
identical since they are not mapped to each other by simply 
reversing the magnetic and the statistical fluxes. In Fig.~\ref{fig:nu2_1}(b), 
a singlet ground state is obtained at $\nu=-2$ with $\theta/\pi=2$. This is 
unique and gives 
$C=-2$, which is consistent with the bosonic IQH 
state~\cite{Senthil_PRL13,Furukawa_PRL13}. Figure~\ref{fig:nu2_1}(b) 
suggests that this is adiabatically connected to the singlet FQH state of
fermions at $\nu=-2/3$~\cite{Wu_CFspin_PRL93}.

\begin{figure}[t!]
 \begin{center}  
  \includegraphics[width=\columnwidth]{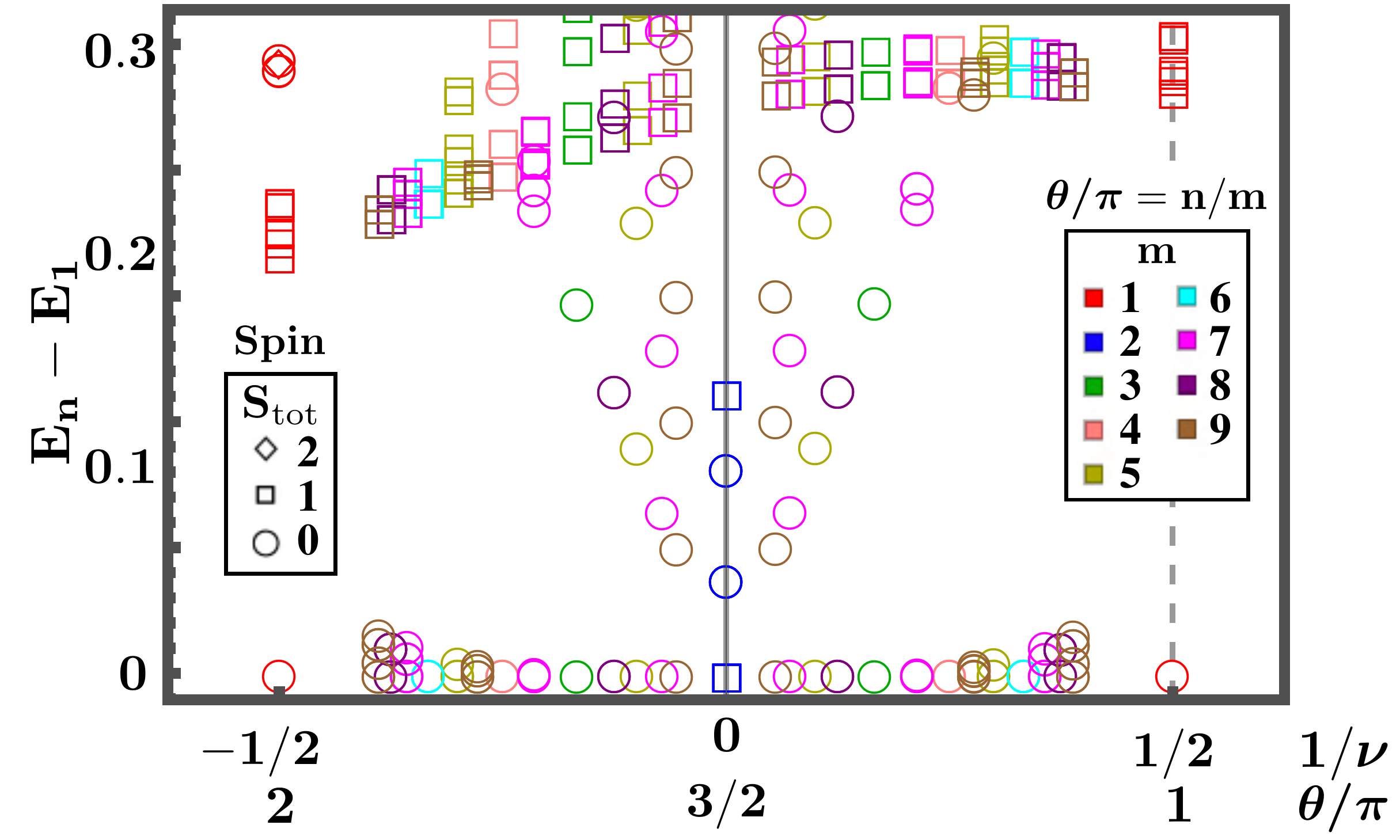}
 \end{center}
 \caption{
 Energy gaps as functions of $1/\nu$ for $V=5$. We set $N_p=6$ and 
 $N_x\times N_y=6\times5$. We plot the lowest 15 energies within
 the $S_z^\text{tot}=0$ sector at each at each $1/\nu$. The vertical dashed 
 lines represents fermionic systems.
 }
 \label{fig:nu2_1_2}
\end{figure}
Let us now focus on the vicinity of $1/\nu=0$ with $\theta/\pi=3/2$ (semions)
in Figs.~\ref{fig:nu2_1}. The energy gap looks symmetric around $1/\nu=0$, and 
closes at $1/\nu=0$. In fact, this symmetry is exact since 
the energy with $S_\text{tot}=0$ should be an even function of $1/\nu$ when
$N_p=4$ (see Appendix~\ref{appx:fourspin}). We then plot the energy gap but for
$N_p=6$ in Fig.~\ref{fig:nu2_1_2}. Even though the systems no longer have 
the emergent symmetry, the energy gaps with $S_\text{tot}=0$ at 
$\pm\nu$ are almost symmetric (e.g. the gap of the first excited states 
$\Delta E(1/\nu)$ at $1/\nu=\pm1/6$ gives 
$\Delta E(1/6)-\Delta E(-1/6)\approx0.0008$). This symmetric behavior is 
consistent with the paring of two 
semions~\cite{Laughlin_anyonS_PRL88,Fetter_anyonS_PRB89,CHEN_anyonS_B89,Hasegawa_fluxS_PRB90,Hosotani_PRB90,Lee_PRL90,Balatsky_PRB91}
since the sign of fluctuations of
$1/\nu$ has no influence for bosons. The gap closing at $1/\nu=0$ is also
consistent with the emergence of the Nambu-Goldston modes of the spinful anyon 
superconductor.

\subsubsection{Topological degeneracy and Chern number}
\begin{figure}[t!]
 \begin{center}  
  \includegraphics[width=\columnwidth]{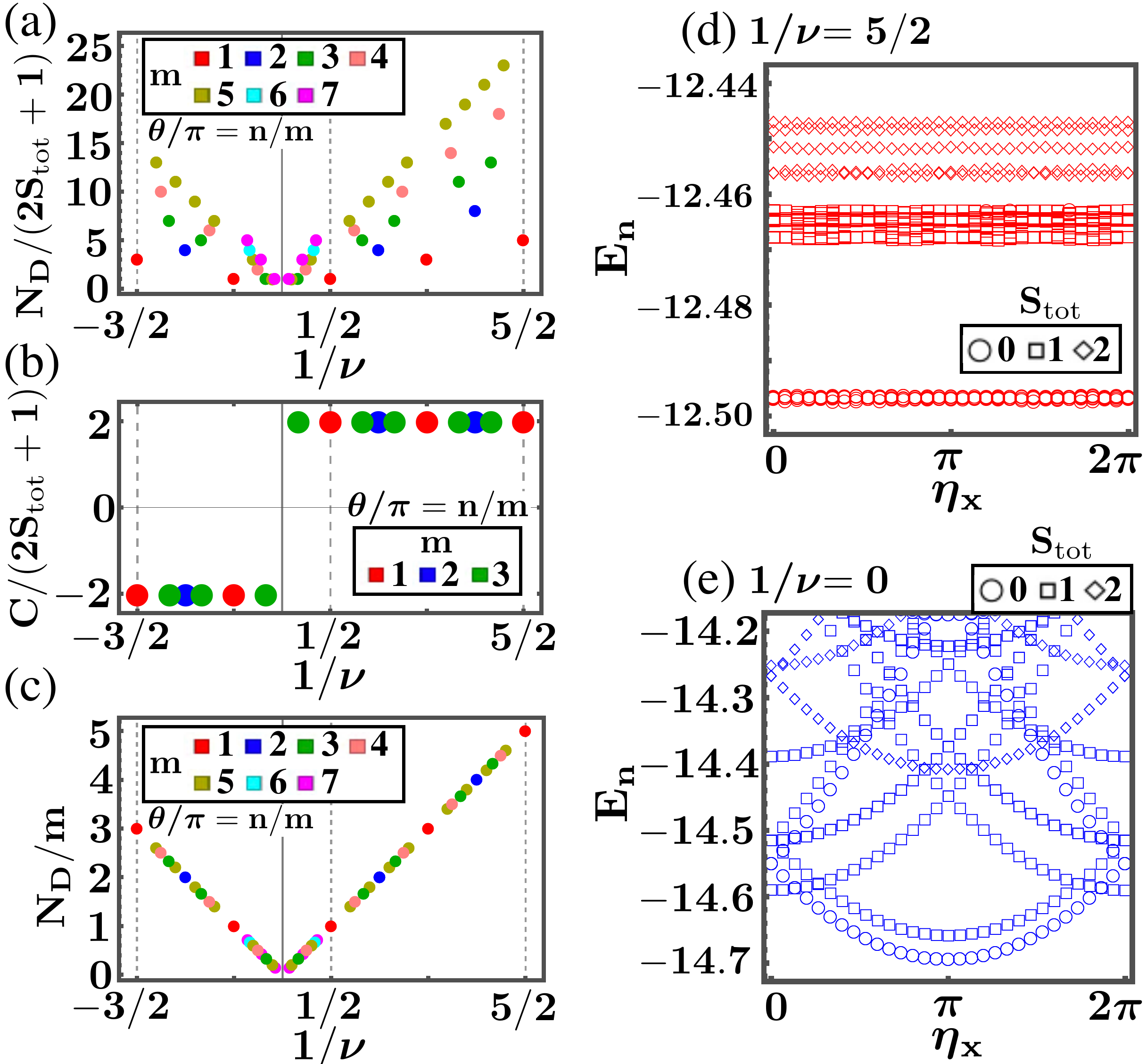}
 \end{center}
 \caption{
 (a) Ground state degeneracy $N_D$ divided by the spin degeneracy 
 $2S_\text{tot}+1$ ($=1$ in this case) as a function of $1/\nu$.
 (b) Many-body Chern number.
 (c) Ground state degeneracy $N_D$ divided by the denominator of $\theta/\pi$. 
 (d)(e) Spectral flows at (d) $1/\nu=5/2$ of fermions and (e) 
 $1/\nu=0$ with $\theta/\pi=3/2$.
 In the all panels, the same setting as Fig.~\ref{fig:nu2_1}(b) is used.
 }
 \label{fig:nu2_2}
\end{figure}
Let us discuss topological properties of the ground states. Assuming that 
states are degenerate if their energy difference is less than 0.001 in
Fig.~\ref{fig:nu2_1}(b), we plot 
the ground state degeneracy $N_D$ and their 
many-body Chern number $C$ in Figs.~\ref{fig:nu2_2}(a) and 
\ref{fig:nu2_2}(b), respectively. Even though $N_D$ discretely changes as
$\nu$ is changed, $C$ remains constant and its sign changes at $1/\nu=0$. This 
suggests that each gap is characterized by the many-body Chern number $C$.

Validity of the the generalized St\v{r}eda formula in Eq.~\eqref{eq:GSF} is 
nontrivial for the singlet ground states unlike the spin-polarized case. 
In Fig.~\ref{fig:nu2_2}(c), we plot $N_D/m$ as a function of $1/\nu$. The slop 
is identical to $C$ ($=\pm2$ in this case), which is actually consistent with
Eq.~\eqref{eq:GSF}. In fact, Eq.~\eqref{eq:GSF} generally holds in the spinful 
systems by assuming some conditions as derived below.

For simplicity, we consider a translational invariant system as discussed in 
Ref.~\onlinecite{Kudo_PRB20}. We define the magnetic translational operators 
$\tau_{i\alpha}$ and 
$\rho_{i\alpha}$ for $i$th fermions (with statistical fluxes) with spin 
$\alpha=\uparrow,\downarrow$ along noncontractible loops on the torus in the 
$x$ and $y$ directions, respectively. They satisfy
\begin{align}
 \rho_{i\alpha}^{-1}\tau_{j\beta}\rho_{i\alpha}\tau_{j\beta}^{-1}=e^{i2\theta},
\end{align}
for any $\alpha,\beta$ since the left-hand side is transformed to a local move 
of the particle $i$ around the particle 
$j$~\cite{Birman_Braid_M69,Einarsson_PRL90,Wen_anyon_PRB90}. This implies 
\begin{align}
 [\tau_{i\alpha}^m,\rho_{j\beta}]=0
\end{align}
for $\theta/\pi=n/m$, meaning that the Hamiltonian specified by twisted 
boundary conditions should commute with $\tau_{i\alpha}^m$ and $\rho_{j\beta}$.
Then defining the translation operators of center of 
mass~\cite{Haldane_TD_PRL85} in the same way of Ref.~\onlinecite{Kudo_PRB20}, 
one obtains at least $pm/|\nu|$-fold degeneracy at $\nu$ with $\theta/\pi=n/m$
in a family of $\nu=p$ IQH system. This reduces to 
\begin{align}
 N_D=Cm/\nu,
\end{align}
by assuming that the ground state does not have any other degeneracy and gives 
the Chern number as $C=\text{sgn}\{\nu\}\times p$ as shown in 
Fig.~\ref{fig:nu2_2}(b). Taking its difference, we obtain Eq.~\eqref{eq:GSF}. 

In Figs.~\ref{fig:nu2_2}(d) and \ref{fig:nu2_2}(e), we plot the energies as 
functions of $\eta_x$ with $\eta_y=0$ at $1/\nu=5/2$ and $1/\nu=0$. Here the 
systems in Fig.~\ref{fig:nu2_1}(b) are used. While the FQH state at $\nu=2/5$ 
is nearly independent of the boundary condition, the spectral flow at $1/\nu=0$
has the strong $\eta_x$-dependence and the gap is closed. This is consistent 
with the emergence of the Nambu-Goldston modes of the spinful anyon 
superconductor. 

\section{Conclusion}
In this paper, the extended Hubbard model of anyons are numerically analyzed. 
In a family of $\nu=1$ IQH system, we have confirmed the maximally 
polarized ground states during the evolution of the flux-attachment. In a 
family of $\nu=2$ IQH system, we have shown that the singlet IQH state at 
$\nu=2$ is adiabatically connected to the $\nu=2/5$ singlet FQH state. The 
adiabatic continuity between the bosonic IQH state at $\nu=2$ and the singlet 
FQH at $\nu=2/3$ is also confirmed. These results are consistent with the 
composite fermion theory with spin~\cite{Wu_CFspin_PRL93}. The many-body Chern 
number not only works as an adiabatic invariant, but also describes the 
wild change of the topological degeneracy during the evolution.

The generalized St\v{r}eda formula~\cite{Kudo_PRB20} for spinful anyons are 
proposed and its validity is confirmed in various QH systems. In adiabatic 
evolution, what is deformed continuously is not states themselves but a gap 
between the sets of the multiplets.

\acknowledgements
We thank Jainendra K. Jain for helpful comments on the mechanism of the gap 
closing in the $\nu=1/2$ bosonic system. We also thank Tomonari Mizoguchi for 
useful discussions on the singlet states of four spins. 
We thank the Supercomputer Center, the Institute for Solid State Physics, the 
University of Tokyo for the use of the facilities.
The work is supported in part by JSPS KAKENHI Grant No.~JP17H06138,
the JSPS Fellowship for Research Abroad (K.K.), and JST CREST Grant 
No.~JPMJCR19T1.

\appendix
\section{Emergent symmetry of four-particle system}
\label{appx:fourspin}
Let us consider the system with the statistical phase $\theta$ at the filling 
factor $\nu$. Since this system is mapped to that with $(-\nu,-\theta)$ by 
reversing the magnetic and the statistical fluxes, the energy satisfies
\begin{align}
 E(\nu,\theta)=E(-\nu,-\theta).
 \label{eq:Ene1}
\end{align}
In the four-particle system, there is another constraint of $E(\nu,\theta)$. 
The eigenstates of $\bm{S}_\text{tot}^2$ and $S_z^\text{tot}$ with $S_\text{tot}=S_z^\text{tot}=0$ are doubly degenerate~\cite{Tsunetsugu_JPSJ01}:
\begin{align}
 \ket{\pm}=[(12)+e^{\pm i\phi}(13)+e^{\mp i\phi}(14)]/\sqrt{3},
\end{align}
with $(ij)\equiv(S_i^-S_j^-\ket{\uparrow\uparrow\uparrow\uparrow}
+S_i^+S_j^+\ket{\downarrow\downarrow\downarrow\downarrow})/\sqrt{2}$
and $\phi=2\pi/3$.
They satisfy 
\begin{align}
P_{34}\ket{\pm}=\ket{\mp}
 \label{eq:P34_1},
\end{align}
where $P_{34}$ is the exchange operator between the spins 3 and 4. This implies
the following equivalence:
\begin{align}
 \Phi^\dagger P_{34}\Phi\simeq\text{diag}\{1,-1\},
 \label{eq:P34_2}
\end{align}
where $\Phi=(\ket{+},\ket{-})$. Equations~\eqref{eq:P34_1} and \eqref{eq:P34_2}
give
\begin{align}
 E_{S_\text{tot}=0}(\nu,\theta)=E_{S_\text{tot}=0}(\nu,\theta+s\pi),
 \label{eq:Ene2}
\end{align}
with $s$ integer.

According to Eq.~\eqref{eq:AH}, a family of $\nu=p$ IQH system gives the
following constraint between $\nu$ and $\theta$:
\begin{align}
 \theta=\theta(\nu)\equiv\pi\left(\frac{p+1}{p}-\frac{1}{\nu}\right)
\end{align}
This satisfies
\begin{align}
 \theta(-\nu)=-\theta(\nu)+\frac{2(p+1)}{p}\pi.
 \label{eq:theta}
\end{align}
With $p=2$, we have 
\begin{align}
 E_{S_\text{tot}=0}(-\nu,\theta(-\nu))
 =E_{S_\text{tot}=0}(\nu,\theta(\nu)),
\end{align}
where Eqs.~\eqref{eq:Ene1} and \eqref{eq:Ene2} are used. This implies that the 
energy with $S_\text{tot}=0$ is an even function of $1/\nu$.

\bibliographystyle{apsrev4-2}
\bibliography{citation}

\begin{thebibliography}{80}
\bibitem [{fer()}]{fermi}%
  \BibitemOpen
  \href@noop {} {}\bibinfo {note} {The Hamiltonian is also expressed by using
  the creation-annihilation operators of hard-core bosons with the statistical
  flux. In the numerical calculations, we diagonalize it because of its
  technical simplicity.}\BibitemShut {Stop}%
\bibitem [{cal()}]{calS}%
  \BibitemOpen
  \href@noop {} {}\bibinfo {note} {We identify $S_\text{tot}$ by counting the
  degeneracy, assuming that states are degenerate if the energy difference is
  less than 0.001. The energy is computed by diagonalizing the
  block-diagonalized Hamiltonians specified by $S_\text{tot}^z$ with the
  Lanczos method. The Lanczos iteration is finished if the lowest eigenvalue
  $E(i)$ at $i$th step satisfies $|E(i)-E(i-1)|<10^{-6}$. An $n$th excited
  state is computed as the ground state of the matrix in which the lowest $n-1$
  states are shifted energetically. In Fig.~\ref{fig:nu2_2}(d), for example,
  one finds some states with $S_\text{tot}=0$ at $E_n\approx-12.462$ even
  though almost all states around them give $S_\text{tot}=1$. They are expected
  to be numerical errors.}\BibitemShut {Stop}%
\end{thebibliography}

\end{document}